**Vectorial Fluorescence Emission from Microsphere Coupled to Gold Mirror**

*Adarsh B. Vasista, Sunny Tiwari, Deepak K. Sharma, Shailendra K. Chaubey, and G. V. Pavan Kumar[*]*

Adarsh B. Vasista, Sunny Tiwari, Deepak K. Sharma, Shailendra K. Chaubey, and Prof. Dr. G. V.
Pavan Kumar
Department of Physics, Indian Institute of Science Education and Research, Pune-411008, India
E-mail: pavan@iiserpune.ac.in

Prof. Dr. G. V. Pavan Kumar
Center for Energy Science, Indian Institute of Science Education and Research, Pune-411008, India

Keywords: Whispering gallery modes, Momentum space microscopy, Vectorial polarization

We report on the generation, and momentum space distribution of fluorescence emission from individual $SiO_2$ microsphere on dye coated Au mirror. The molecular fluorescence emission mediated via whispering gallery modes of the sphere is studied using polarization resolved optical energy-momentum micro-spectroscopy. Our experiments reveal intensity dependence of split modes of the cavity as a function of in-plane wavevector and emission polarization in the far field. The exotic far-field distribution can be understood by sphere-image sphere model that further reveals the polarization dependence of the split modes. The presented results reveal the potential of metallo-dielectric soft micro-cavities to engineer molecular emission that can encode spin and orbital angular momentum states and can be further extrapolated to realize dye-loaded active meta-atoms and meta-surfaces.

1. **Introduction**

Engineering molecular emission from micro and nanocavities has implications not only in designing efficient chip-scale molecular emitters,[1-4] but also in understanding fundamental physical and chemical concepts in molecular electrodynamics.[5-7] Specifically, inelastic radiative processes such as molecular fluorescence [5, 8-11] and vibrational Raman scattering [12-14] has been extensively studied to gain deeper insight into how molecule behaves when they interact with an optical cavity. To this end, a variety of cavity architectures have been recently explored to control light-molecule interactions including plasmonic micro and nano-cavities, both in strong [6, 8, 15] and weak coupling[16] regime. On the other hand, dielectric microcavities have a rich history, [17-19] particularly in the context of molecule-cavity interaction. [20-22] By combining plasmonic and dielectric structures, hybrid metallo-dielectric cavities have been realized. [23-26]

Hybrid plasmonic-photonic cavities have shown to manipulate morphology-dependent resonances, [27, 28] engineer light-matter interactions, [29, 30] detect molecules, [31, 32] etc. In essence, such hybrid cavities can potentially bring together the *'best of both worlds,'* and there is an imperative to study them in detail, especially in the context of understanding how molecules emit in such cavities. In this context, interesting questions to ask are*: how is the fluorescence of molecule influenced in such hybrid cavities? What are the spectral, wavevector and polarization features of far-field molecular fluorescence in such cavities?* Addressing them can lead to some unexpected insights into molecule-cavity interaction, and can potentially lead to some important applications including orbital angular momentum carrying molecular light sources and chiro-optical sensors.

Motivated by these issues, herein we report our experimental studies on in-plane momentum and polarization-resolved fluorescence emission from dye molecules coupled to microsphere on a gold mirror. One of the important findings from our study is that the far-field fluorescence emission from the cavity is vectorial. That is, the far-field fluorescence spectra have strong polarization dependence in radial and azimuthal coordinates. To the best of our knowledge, this is a unique and unreported observation in molecular fluorescence from an optical cavity.

### 1.1. Background

Strong coupling of microresonators results in the splitting of individual whispering gallery modes (WGMs) into bonding and antibonding modes and can show rich polarization structures and radiation patterns. Since WGMs can carry azimuthal polarization, [33-35] strong coupling in these structures can potentially open up new areas of research by imparting exotic polarization signatures to molecular emission coupled to the system.

In the past, strongly coupled microresonator system has been realized by placing microspheres side-by-side. [36, 37] Strong coupling in microresonators has also been studied in semiconductor microdiscs, [38, 39] microspheres,[40, 41] epitaxial heterostructures [42] etc. One way to create a strongly coupled microresonator system is to utilize sphere-image sphere interaction by placing microsphere on a metallic mirror. This configuration has been utilized with metal/dielectric nanoparticles and nanowires in various prospects, [12, 15, 25, 43, 44] but has not been studied extensively in understanding polarization structure of strong coupling physics. This configuration is advantageous over in-plane coupling of microspheres as the size mismatch problem between microspheres is solved.

### 2. Architecture of the system:

**Figure 1** (a) shows the schematic of the experiment. A gold mirror of 100 nm thickness was deposited using thermal vapor deposition method on a glass coverslip. Nile blue (NB) molecules ($10^{-4}$M) were dropcasted on the metal mirror and allowed to dry to form a thin molecular film. The number of molecular layers in the focal volume is more than one for this concentration, and hence molecular fluorescence from the layers not in contact with the metal mirror will be screened and not quenched strongly. SiO$_2$ microspheres of size ~3 µm (with 5% variation in diameter) were dropcasted over the molecular film. This experimental configuration is an important one as this can be easily extrapolated to study molecular emissions coupled to the WGMs of the structure both in reflection geometry and transmission geometry. An individual microsphere was excited by focusing 633 nm laser beam using a high numerical aperture objective lens. Microsphere acts as a micro-lens and focusses the incoming laser to a very small region through photonic nanojet effect. [45, 46] This creates a large electric field at the junction of the microsphere and the metal mirror. NB molecular fluorescence was enhanced due to strong localization of the electric field. The molecular fluorescence couples to WGMs of the microsphere and emits in specific directions. Inset of Figure 1 (a) shows the brightfield optical image of microsphere on thin Au mirror. Individual microspheres were excited using 100x, 0.95 NA air immersion objective lens and the backscattered light was collected from the same objective lens. The collected light was projected onto the Fourier plane (FP) and then analyzed for spectroscopic information (see section S1 of supplementary information for experimental setup).

## 3. Results and discussions

The NB fluorescence spectrum collected from individual SiO$_2$ microsphere is shown in figure 1 (b). One can see the enhancement of WGM coupled NB fluorescence from the microsphere coupled Au mirror when compared to that on the glass substrate ($I_{gold}/I_{glass}$ ~ 5 with Q of individual split mode ~300). This enhancement is due to very high electric field generated at the metal-microsphere interface which enhances the WGM coupled molecular fluorescence (see section S2 and S3 of supplementary information for molecular fluorescence spectra from the bare glass substrate, Au mirror, electric field distribution at the microsphere-substrate interface, and decay rate respectively.) Also the WGMs show clear splitting when the microsphere is coupled to Au mirror,[47] as shown in figure 1 (c) (energy of splitting: 8.27 meV).

**3.1. Fluorescence emission via Mie modes in the geometry**

To understand and identify the Mie modes, the molecular fluorescence spectrum was collected by placing microsphere on the glass substrate (see figure 1 (c)). Mie modes were characterized using radial ($n$), orbital ($l$) and azimuthal ($m$) quantum numbers (see section S4 and S5 of supplementary information for discussion on assignment of Mie modes). For an isolated microsphere in vacuum, azimuthal quantum number ($m$) has *2l+1* degeneracy. This degeneracy can be lifted by coupling it to another microsphere, due to the presence of material inhomogeneity or a scatterer at the periphery of the resonator,[48-50] deviation of the morphology of the microresonator from that of perfect sphere, etc. The lifting of degeneracy will be manifested as a spectral split in the Mie modes. In the present case, the splitting is mainly due to the presence of a metal mirror which alters the azimuthal symmetry of the microsphere. Figure 1 (c) clearly shows the Mie mode $TE_{16,1}$ is split into two modes when microsphere is placed on Au mirror.

### 3.2. Optical energy-momentum spectroscopy of fluorescence emission from the geometry

To analyze the momentum space signatures of WGMs, the outcoupled fluorescence from the system was projected onto the Fourier plane. **Figure** 2(a) shows the FP image of the captured light from the microsphere coupled to Au mirror. The FP image clearly shows that the emission is ring-shaped. To understand the effect of coupling to the metal mirror, we measured FP image from microsphere coupled to NB coated glass substrate. Figure 2 (b) shows the measured FP image. One can note that the emission of microsphere on glass occupies a relatively broad range of wavevectors in comparison to microsphere coupled to the gold mirror (see section S6 of supplementary information for quantification of FP images). This typical emission pattern is due to the coupling of molecular fluorescence to WGMs of the microresonator system. To ascertain this, we performed similar experiments on ~1 µm size $SiO_2$ microsphere on Au mirror. At this operational wavelength, WGMs were not excited, but the molecular emission (both fluorescence and Raman) were enhanced. The FP image shows that the emission is uniform over a broad range of wavevectors, unlike the present case (see section S7 of supplementary information for FP image and spectrum).

Note that the FP images shown in figure 2 (a) and (b) are convoluted images of all the excited WGMs in the microsphere. If one filters specific $k/k_0$ values in the FP image and then disperse to provide wavelength resolution, then momentum distribution of individual WGMs can be studied. The energy-momentum (E-k) spectroscopy measurement is very important for microsphere coupled to thin Au mirror, as coupling introduces split in WGMs and the E-k spectrum sheds light on momentum distribution of the split modes in such situation.

With this hindsight we performed E-k spectroscopy on microsphere coupled to NB coated Au mirror and glass substrate. A thin slice of the FP image (see figure 2 (a) and (b)) around $k_x/k_0=0$ was filtered using the slit of the spectrometer and dispersed using 1200 grooves/mm diffraction grating to obtain the E-k spectrum. Figure 2 (c) and (d) show the corresponding E-k spectra for microsphere coupled to Au mirror and glass substrates respectively. The Mie mode split (originally $TE_{16,1}$ for unperturbed microsphere) observed near 660 nm wavelength (peaks 'a' and 'b') clearly shows different momentum distributions (see figure 1 (c)). We observe emission at lower $k_y/k_0$ values for the peak 'a,' while that for peak 'b' at higher values of $k_y/k_0$. The trend is repeatable, and the data shown here is representative in nature. For the microsphere on glass substrate case, there is no split in the WGMs and this translates to the E-k diagram as well. Both modes emit at larger values of $k_y/k_0$ (see figure 2 (d)). Hence, the coupling of microsphere to a thin mirror not only splits the WGMs in the wavelength domain but also in momentum domain, showing anti-crossing. This anti-crossing behavior of Mie modes is a result of the coupling of the microresonators.

**3.3. Vectorial fluorescence emission**

We were interested in understanding the polarization signatures of the split whispering gallery modes due to strong coupling. To this end, we performed polarization resolved FP imaging and spatially filtered spectroscopy. Polarization-resolved FP images did not show any clear evidence of a polarization structure. Also when output polarization signatures of the split modes were measured using real plane spectroscopy, there was no major difference between the intensity of the split modes (See section S8 and S9 of supplementary information for polarization resolved FP images and real plane spectroscopy).

WGMs can potentially possess orbital angular momentum with azimuthal polarization signatures and is difficult to detect its polarization using real plane spectroscopy. Hence, one has to perform polarization resolved E-k spectroscopy to understand the polarization information of strongly coupled optical microresonators. Hence we probed individual WGM using polarization resolved E-k spectroscopy. The FP image was projected onto the slit (width = 200 µm) or a square pin hole (200 µm x 200 µm) placed at the conjugate Fourier plane. The slit (pinhole) was used to filter narrow range of $k_y/k_0$ and $k_x/k_0$ (see section S10 - S12 of supplementary information for E-k spectroscopy using slit and pinhole).

**Figure** 3 shows polarization resolved E-k spectra of individual WGM from strongly coupled microresonators. Square pinholes were placed along $k_x/k_0 = 0$ (figure 3 (a) and (b)) and along $k_y/k_0 = 0$ (figure 3 (c) and (d)) to filter narrow range of wavevectors and the output polarization was analyzed. As seen in figure 2 (b), the split modes outcouple at different wavevectors. So, it is easy to understand them by filtering narrow range of k-vectors where only one of the two split modes exists in the k-space. We try to understand the mode $TE_{16,1}$ when microsphere is placed on Au mirror. As the peak 'b' of the split modes outcouple at higher wavevectors (see figure 3 (c)), we filtered out two points (along $k_y/k_0 = 0$ and $k_x/k_0 = 0$) as shown in figure 3 (a) and (c). At point (i) the polarization of the mode was predominantly along $k_x/k_0 = 0$ (0° in analyzer) line. But at point (ii), the polarization of the mode has rotated by 90°, and the mode is polarized along $k_y/k_0 = 0$ (90° in analyzer). Thus, it shows that the polarization of the Mie mode rotates as one moves along the momentum plane making it an azimuthally polarized beam.

Interestingly, the peak 'a' of the split modes behaves the same (see figure 3 (b) and (d)). But the polarization signatures are orthogonal to that of split mode 'b,' making it radially polarized. Split modes due to strong coupling of microresonators, thus, possesses vectorial polarization signatures and are azimuthally and radially polarized respectively. This result can potentially have wide relevance in understanding the angular momentum mechanics inside strongly coupled cavities. Since both the modes outcouple at different wavevectors, one can easily filter out a specific mode. Under far field real plane spectroscopy/imaging direct observation of vectorial polarized emission is not possible (see section S9 of supplementary information).

**3.4. A model to explain observed results**

We can understand the momentum and polarization signatures of the WGMs by looking at the electromagnetic field of an individual WGM. For simplicity, we consider the electromagnetic field of an individual microsphere. The results can then be extrapolated to microsphere over the mirror. The electromagnetic field of a WGM of an isolated sphere is concentrated at a distance of $r(1 - \sqrt{1 - |m/l|^2})$ from the substrate, where $r$ is the radius of the microsphere[48] (see **figure 4 (a)**). The equatorial plane of the sphere corresponds to $|m| = l$. In non-degenerate case, the states are filled in the order of increasing energy and mode $|m| = l$ will have the lowest energy. Also, one can find that the distance of the mode increases as the mode number $l$ decreases for any $m$. So, modes with lower $l$ value are farther away from the substrate and will be nearer to the equatorial plane. FP image of the light captured from the microsphere has a resolution in terms of wavevectors $k_x$ and $k_y$, which is in turn related to the elevation and azimuthal emission angles. Intensity pattern in the FP image depends on the projection of a mode, $(l,m)$, onto $k_x$ and $k_y$ axes. Modes with different values of $|m|$, for a particular $l$, project onto different $k_x$ and $k_y$ values. Hence there is an intimate relation between the modes generated in the microsphere and the emission FP patterns (see section S14 for a detailed discussion on Mie modes and its Fourier space signatures).

For quasi-TE modes of the microsphere, where the polarization of the electric field is in-planar, the polarization of the light trapped inside the sphere suffers a continual change during total internal reflection (TIR). The state of polarization (SOP) of the WGMs, for TE modes, preserves this cylindrical symmetry of the microsphere and generates cylindrical vector beams. [33] When the microspheres are strongly coupled, the modes sensitive to the coupling will split. [39] To understand the polarization states of the outcoupled radiation, it is imperative to study how the Mie modes are generated.

We modeled the microsphere on metal film using bi-sphere system. The calculated spectra using the bi-sphere model is shown in figure 4 (b). Dipoles with different orientations were placed in between two spheres, and the intensity is plotted with respect to the wavelength of the oscillating dipole. Due to coupling with the second microsphere, once degenerate mode ($TE_{16,1}$) now splits into two different modes (see section S13 of supplementary information). Polarization degeneracy in excitation is lifted and the mode 'a' and 'b' are excited using orthogonal input polarizations (see figure 4 (c) - (e)). Both modes will suffer continuous TIR along the periphery of the sphere. As the input polarization of the excited modes is orthogonal, the outcoupled radiation will also preserve this orthogonality. Since the system is cylindrically symmetric, the polarization of the modes which outcouple after suffering TIR along the periphery of the sphere will be vectorial. This model faithfully explains the momentum space and polarization signatures from strongly coupled optical microresonator system and warrants more rigorous theoretical study.

## 4. Conclusion

In summary, we have experimentally shown how a strongly coupled metallo-dielectric microsphere cavity can influence molecular fluorescence to carry exotic polarization structure in the far-field regime, thus representing vectorial fluorescence emission. Using polarization and momentum resolved optical spectroscopy, we show the split modes of microresonator outcouple with azimuthal and radial polarizations. Thus the system under study is a possible generator of two vector beams with orthogonal polarization, which are distributed differently in momentum space. Given that microsphere can be deposited at the desired location and shrunk in size by methods such as reactive ion etching, our studies can be extrapolated to realize active meta-atom and meta-surfaces.

**Supporting Information**

Supplementary material containing following details is available from Wiley Online Library.

(i) Sample preparation

(ii) Experimental setup

(iii) Results of control experiments

(iv) Details of numerical simulation

**Acknowledgments**

This work was partially funded by DST-Nano mission Grant, Govt. of India (SR/NM/NS-1141/2012(G)) and Center for Energy Science (SR/NM/TP-13/2016), Indo-French Centre for the Promotion of Advanced Research (IFCPAR) (55043), DST-SERB grant (SB/OC/OM-10/2014) and Indian National Science Academy (INSA) grant. A.B.V and S.T thank Vandana Sharma, Dr.Rohit Chikkaraddy and Dr.Vijay Kumar for fruitful discussions. S.T. thanks Infosys foundation for financial aid.

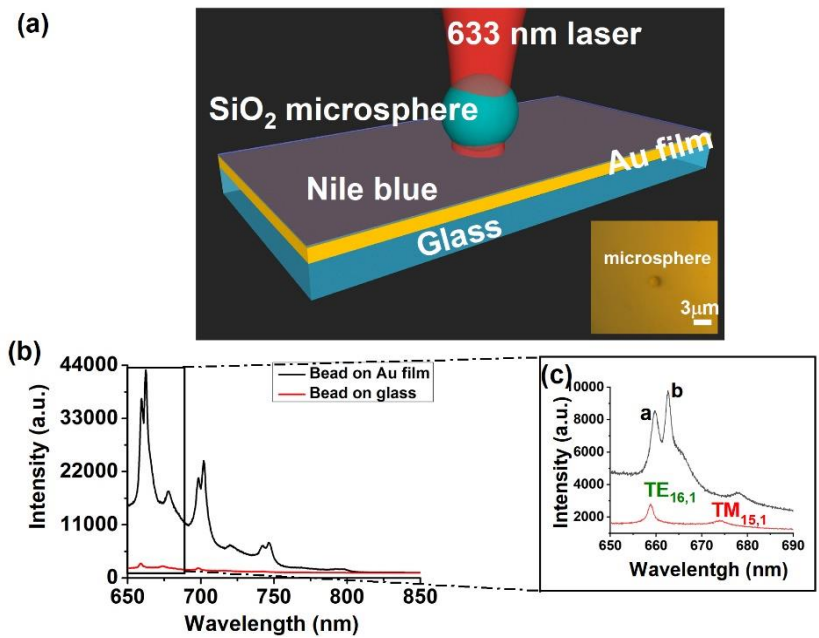

**Figure 1.** (a) Conceptual schematic of metal film coupled $SiO_2$ micro-resonator system. The $SiO_2$ microsphere of size 3 µm, placed on 100 nm thick Au film was excited using a high numerical aperture objective lens. Nile blue molecules were sandwiched between microsphere and Au mirror. *Inset*: Real plane bright-field image of an individual microsphere on gold mirror. (b) Nile blue fluorescence spectra collected by exciting an individual microsphere placed on Au mirror and glass substrate separately. (c) Molecular fluorescence spectrum collected by exciting single microsphere placed on glass substrate and Au mirror captured at higher resolution (1200 grooves/mm). The mode numbers were assigned using analytical theory for single isolated microsphere. The red curve indicates the molecular fluorescence captured by placing microsphere on glass substrate and the black one indicates that on Au mirror.

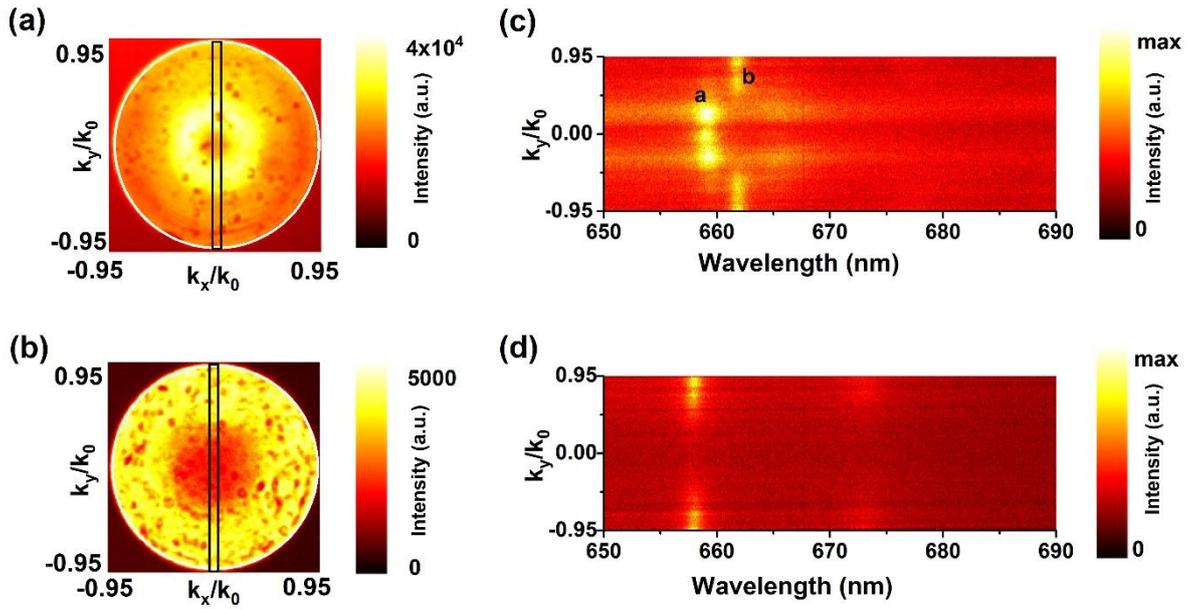

**Figure 2.** (a) Fourier plane fluorescence image captured from microsphere coupled Nile blue coated Au mirror. (b) Fourier space image captured from microsphere coupled Nile blue coated glass substrate showing that emission is ring liked with large wavevector spread. A narrow range of wavevectors around $k_x/k_0$ was filtered (shown as a rectangle in (a) and (b)) and dispersed in wavelength to understand momentum signatures of individual Mie modes. (c) and (d) are measured E-k spectra for microsphere placed on Au mirror and glass substrate respectively.

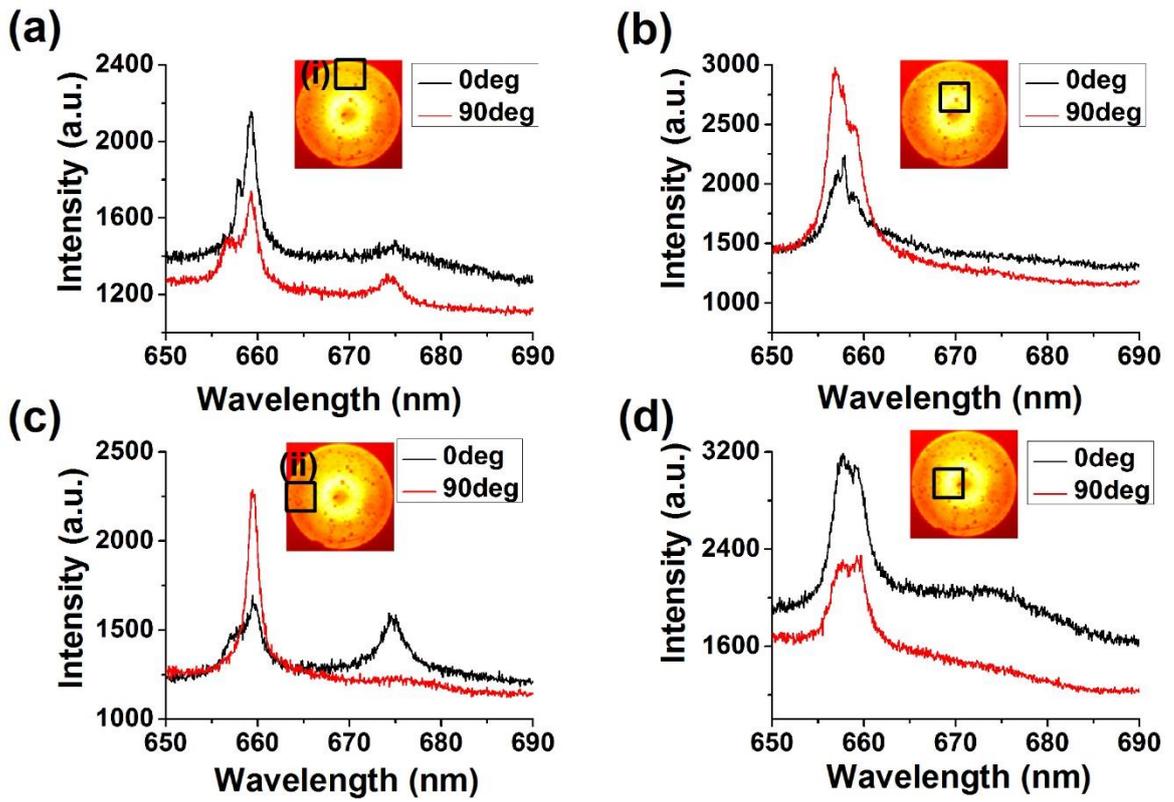

**Figure 3.** Momentum resolved spectra collected from the microsphere coupled to Au mirror system. The out-coupled light at the conjugate back focal plane was filtered using a square pinhole of size 200 µm and was analyzed to understand the polarization signatures at different k-points. (a) and (b) are polarization resolved spectra by placing pinhole at points along $k_x/k_0 = 0$ as shown in the insets. (c) and (d) are polarization resolved spectra by placing pinhole at points along $k_y/k_0 = 0$ as shown in the insets.

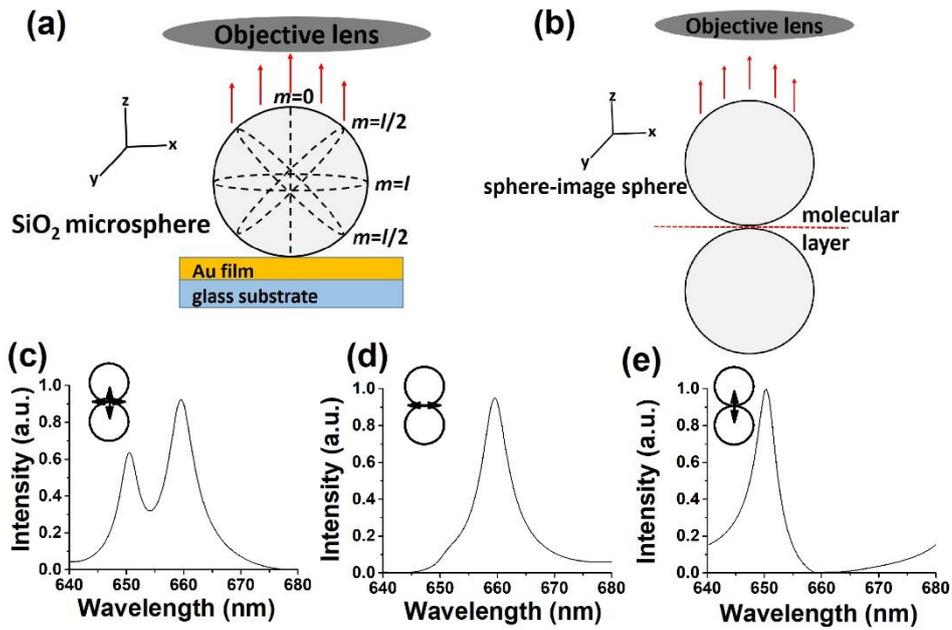

**Figure 4.** (a) Schematic representation of whispering gallery modes in an Au mirror coupled microsphere system. (b) Model system, bi-sphere in contact, considered to represent microsphere on mirror. Calculated spectra for bi-sphere system by placing (c) two orthogonal dipoles (d) single dipole along x-axis (e) single dipole along y-axis.

# Supplementary information

## S1. Experimental setup

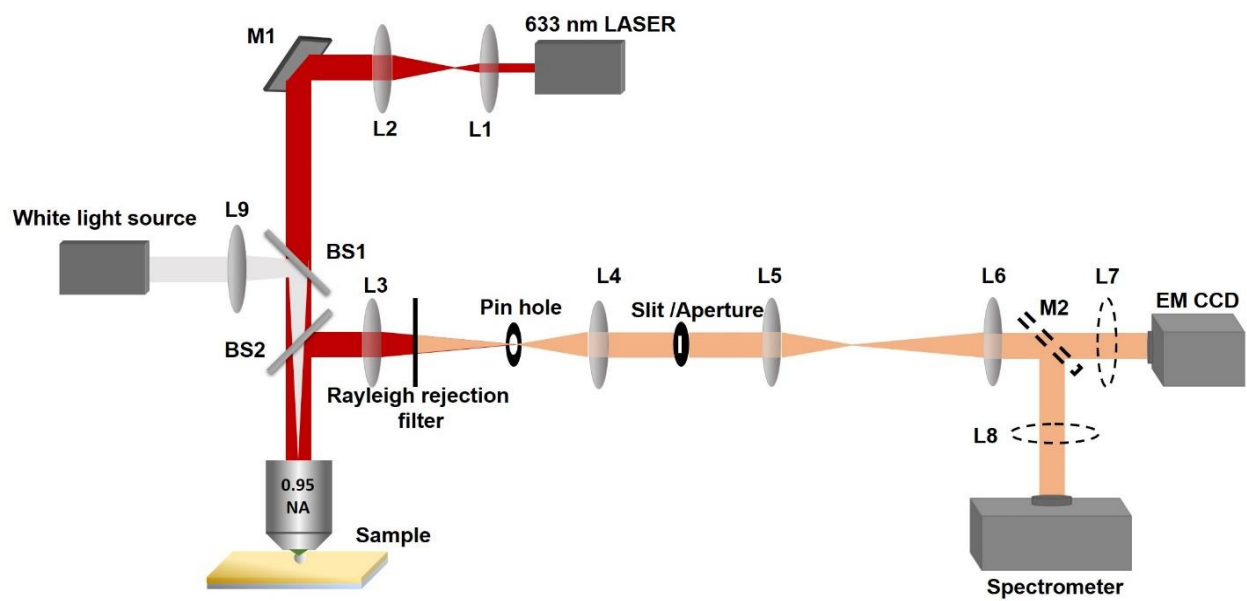

Figure S1: Schematic of the experimental setup used to perform Fourier plane imaging and spectroscopy on microsphere coupled to Au film.

The experimental setup used to probe microsphere coupled to Au film is as shown in fig.S1. Individual microspheres were excited using 100x, 0.95 NA air immersion objective lens and the backscattered light was collected from the same objective lens. The collected light was projected onto the Fourier plane using combination of lenses L3-L6 after rejecting the Rayleigh scattered component using 633 nm Edge filter. Slit /aperture was placed at the conjugate FP created by lens L4 to filter narrow range of wavevectors. Flippable mirror, M2 , was used to direct light to the spectrometer. Lenses L7 and L8 were used to focus the light to the EMCDD and to the spectrometer respectively.

## S2. Molecular fluorescence spectra from bare substrates

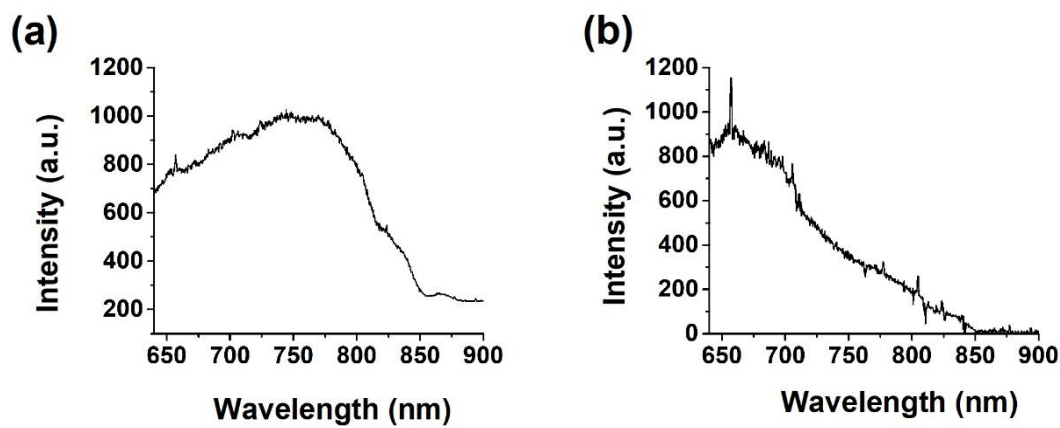

Figure S2: Nile blue molecular fluorescence spectra collected by exciting dye coated (a) glass substrate and (b) Au film respectively.

In the experimental setup, we have utilized high molecular concentration ($10^{-4}$M) by dropcasting dye solution on Au film. Because of this, the number of molecular layers on the film is more than one. Because of this, the molecular layers touching the Au film will be quenched and the layers above it will be screened (see fig.S2). The molecular fluorescence which gets coupled to the WGM will be enhanced and gets outcoupled. On the other hand, Raman can only get enhanced if the frequency overlaps with one of the WGM modes of the sphere. But in any case, fluorescence enhancement always occurs.

## S3. FEM simulations to calculate near field electric field and decay rate

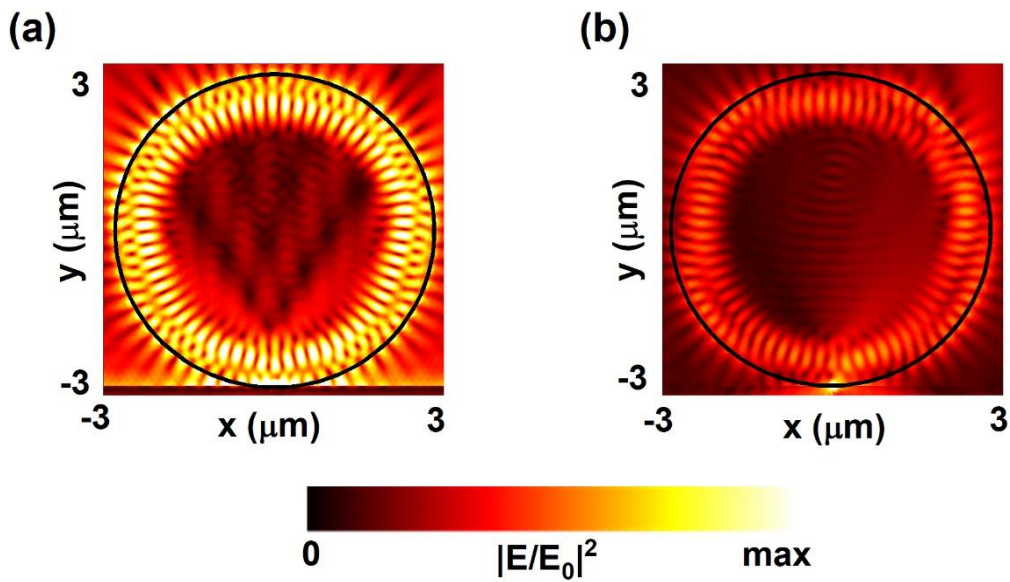

Figure S3: Calculated near field electric field distribution for (a) microsphere on Au film and (b) glass substrate respectively.

To understand the effect of the coupling of microsphere to Au film, we performed finite element method (FEM) based numerical simulations using COMSOL 5.2 software. The size of the microsphere was fixed at 3 µm and was placed on a 100 nm Au film. The simulation area was meshed using free tetrahedral mesh and was terminated using perfectly matched layers (PMLs) to minimize spurious reflections from the boundaries. Wavelength dependent refractive index of gold was taken from [1]. The microsphere was excited using dipole source placed at the substrate – microsphere interface.

Fig. S3 (a) and (b) shows the near field electric field distributions for microsphere placed on Au film and glass substrate respectively. Figure clearly shows enhancement of local electric field for the microsphere on film geometry. This enhancement of the electric field will be manifested in the enhancement of the molecular signatures coupled to the system.

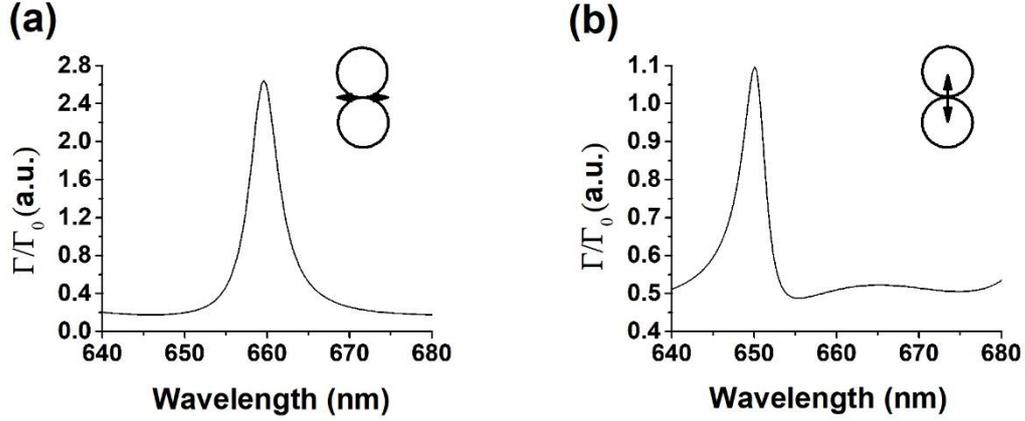

Figure S4: Calculated decay rate for bisphere system by placing dipole oriented along (a) x-axis and (b) y axis respectively.

To understand the decay rate modification by coupling the molecule to microsphere on Au film, we calculated decay rates for individual dipoles using 2D FEM methodologies (COMSOL). Microsphere over metallic film was modeled as bisphere system (sphere-image sphere) and the dipolar source was placed at the junction of spheres. The power radiated by the bisphere was integrated over a circle 4 µm away from the system and normalized by power radiated by the individual dipole. The radiative decay rate is enhanced for dipolar orientation along x-axis as shown in figure S4.

## S4. Assigment of Mie modes

The electromagnetic modes constituted by circular dielectric structures are known as Whispering Gallery Modes (WGMs). The electromagnetic modes get strongly confined due to the suffered total internal reflection of the waves at the surface. Due to the small mode volume and high local electromagnetic field, the Q-value possessed by the microspheres is extremely high [2, 3]. The WGMs of the microsphere, are computed using classical Mie theory and are labelled by two polarizations, Transverse Magnetic (TM) and Transverse Electric (TE) and three quantum numbers $n, l$ and $m$. $n$ is the radial quantum number and corresponds to the number of intensity maxima along the radius of the sphere. $l$ is equal to half the intensity maxima along the great perimeter of the sphere and is termed as the orbital quantum number whereas $m$, which is the azimuthal quantum number, is the projection of $l$ on the quantization axis. The mode number were assigned using the Matlab code provided by [4]. The extinction crossection for a microsphere interacting with an electromagnetic field is given by,

$$Q_{ext} = \frac{2}{(ka)^2} \sum_{n=1}^{\infty} (2n+1) Re(a_n + b_n)$$

where $k$ is the wavenumber and $a_n$ and $b_n$ are mie coefficients to calculate the amplitude of the scattered fields. When $b_n$ is zero, $a_n$ dominates and modes corresponding to this are assigned as TM modes. Similarly, $b_n$ dominates when $a_n$ is set to zero and these modes are assigned as TE modes. Furthermore, the assignment was verified using finite difference time domain (FDTD) simulations.

## S5. Molecular fluorescence spectra

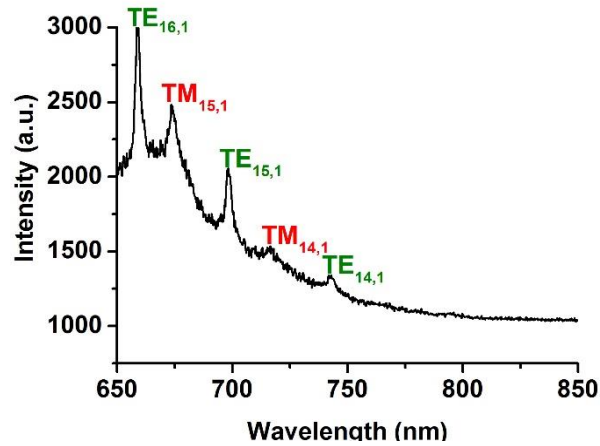

Figure S5: Molecular fluorescence spectra collected from microsphere coupled to glass substrate with mode assignment.

Figure S5 shows the molecular fluorescence spectrum collected by placing microsphere on glass substrate. Individual peaks of the spectrum corresponds to different mie modes of the microsphere. The mode numbers were assigned using analytical mie theory outlined in the previous section S3.

## S6. Quantifying Fourier plane images

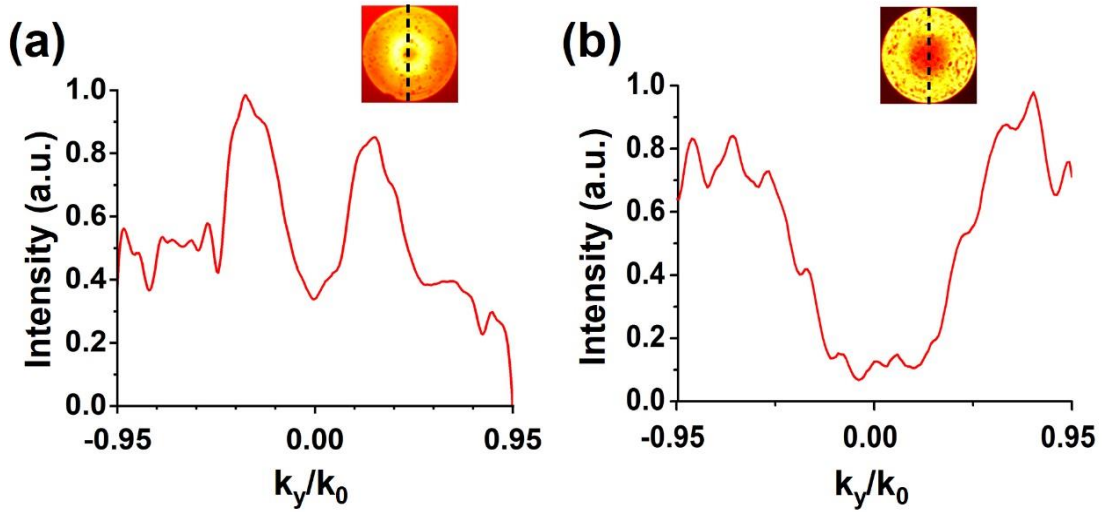

Figure S6: Intensity profiles along $k_x/k_0=0$ line for Fourier plane images captured from microsphere coupled to (a) Au film (b) glass substrate respectively. Insets show Fourier plane images

To further quantify the emission, we plot the intensity of fluorescence emission along $k_x/k_0=0$ and is shown in fig. S6. For microsphere on Au mirror majority of emission happens at $k_y/k_0=$ -0.33 (F.W.H.M=0.32) and 0.29 (F.W.H.M=0.32). And for the microsphere on glass substrate the intensity of emission across $k_x/k_0=0$, fig. S6 (b), shows that emission maximum is at $k_y/k_0$=-0.67 (F.W.H.M=0.55) and 0.75 (F.W.H.M=0.64). Hence coupling to the Au mirror not only affects the spectral shape and causes the spectral split but also affects the emission wavevectors of the WGMs both in terms of emission wavevector maxima and wavevector spread.

## S7. Fourier plane image and spectrum of 1µm beads over Au film

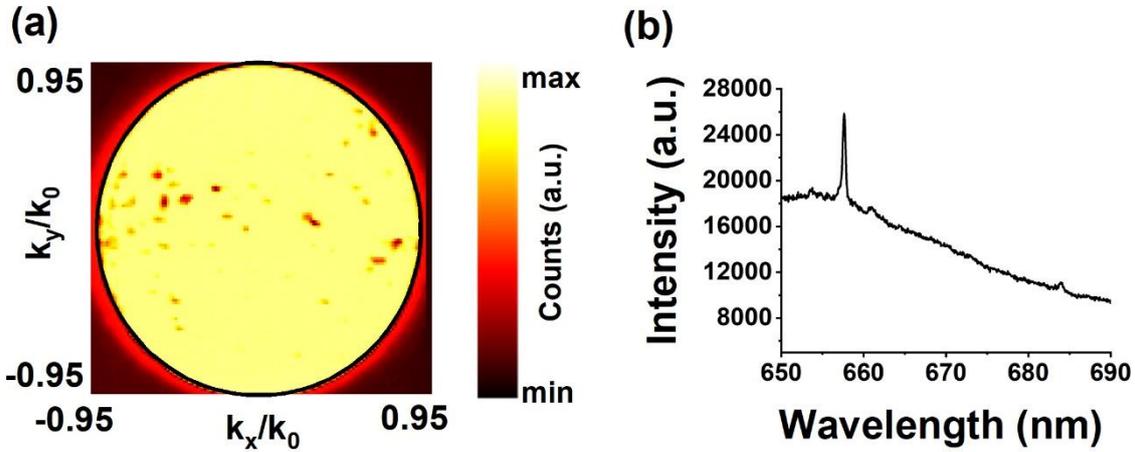

Figure S7: (a) Fourier plane image of an individual 1 µm size $SiO_2$ bead on nile blue coated Au film, showing uniform emission. (b) Nile blue molecular emission spectrum from 1 µm size $SiO_2$ bead on Au film.

Figure S7 (a) denotes the Fourier plane image of the molecular emission captured by coupling it to microsphere of size1µm on Au film. FP image clearly shows that emission is uniform over a broad range of wavevectors. Also, since the molecular emission does not excite WGMs in the microsphere (see figure S7 (b)).

## S8. Polarization resolved back focal plane imaging

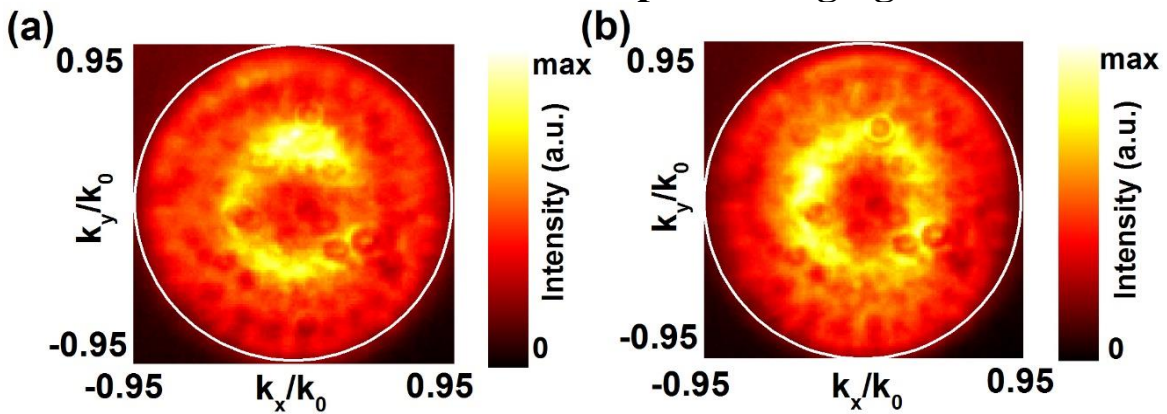

Figure S8: Output polarization resolved fluorescence FP images for microsphere on Au film along (a) $k_x/k_0$ and (b) $k_y/k_0$ axes respectively.

Output polarization resolved FP images for molecular emission along $k_x/k_0$ and $k_y/k_0$ are shown in figures S8 (a) and (b) respectively. Polarization resolved FP images do not provide simultaneous information of momentum and polarization of split modes as the wavelength information is integrated out. This makes the measurement using optical energy momentum diagram imperative.

## S9. Polarization resolved real plane spectroscopy

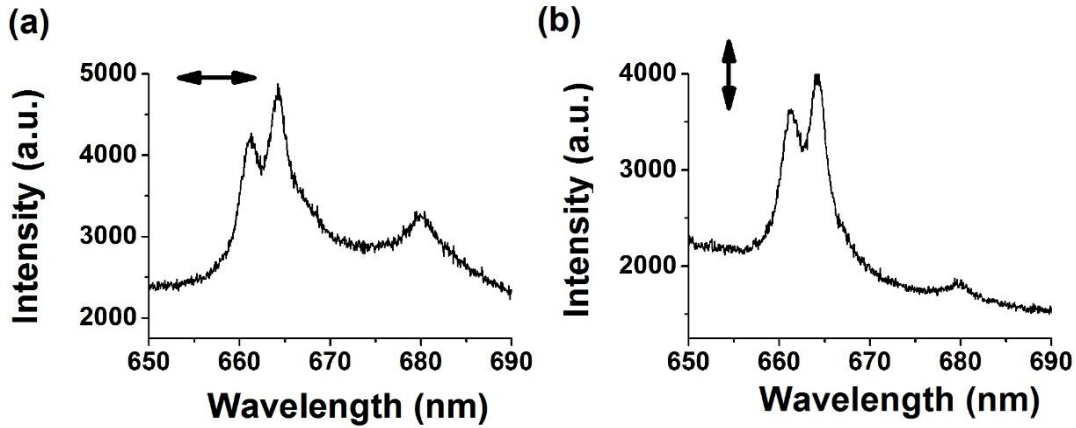

Figure S9: Output polarization resolved spectra collected from strongly coupled optical microresonator system in real plane for (a) 0° and (b) 90° analyzer positions respectively. The spectra show no significant differences between two different analyzer positions.

## S10. Momentum resolved spectroscopy of split mie modes

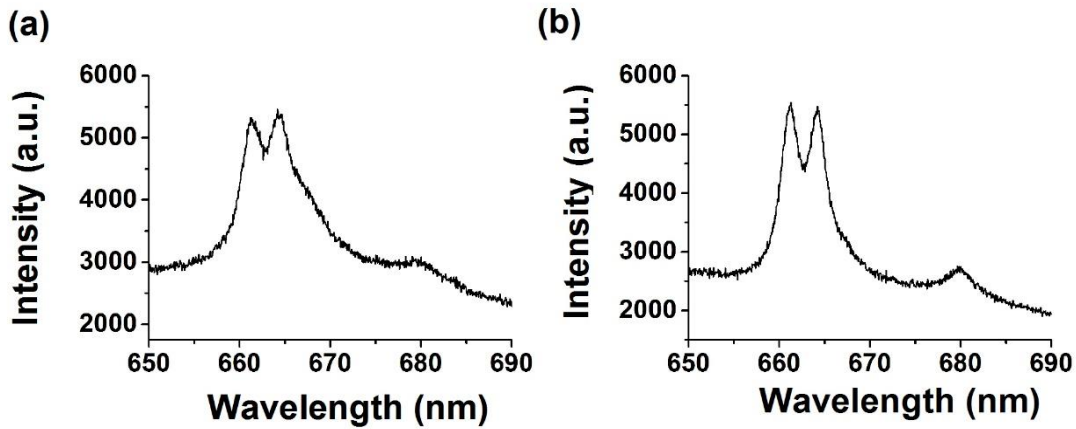

Figure S10: Momentum resolved spectra collected from strongly coupled optical microresonator system by placing a 200 µm slit in the image plane along (a) $k_x/k_0 = 0$ and (b) $k_y/k_0 = 0$.

# S11. Polarization resolved E-k spectroscopy using a slit

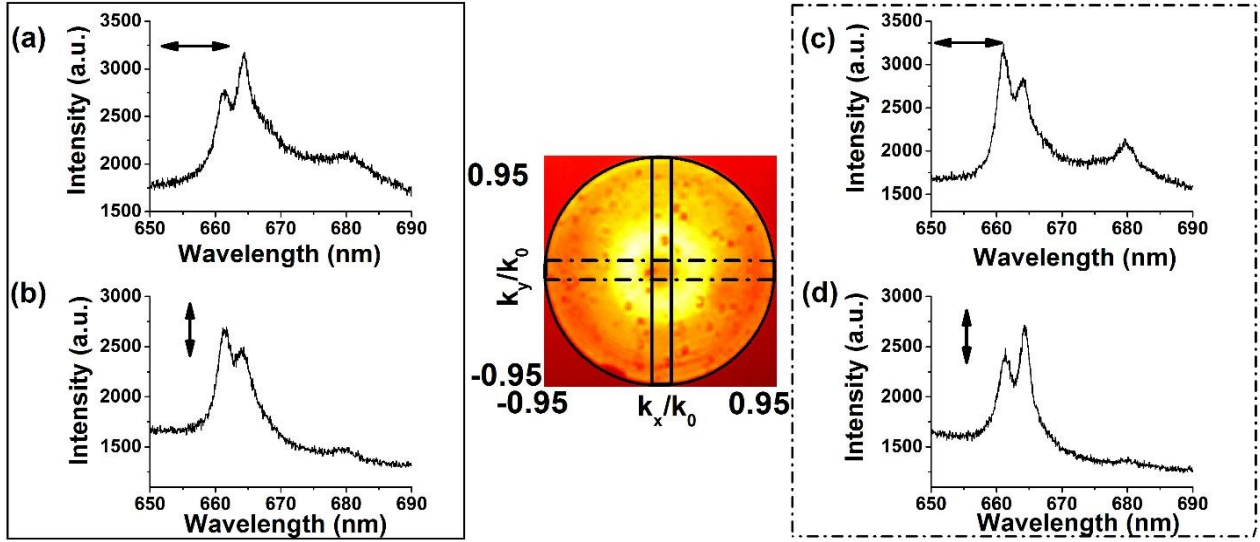

Figure S11: Momentum resolved spectroscopy of individual WGMs. (center) Fourier plane image of emission from microsphere coupled to nile blue film coated Au film. Momentum resolved spectroscopy was performed by filtering out a narrow range of wavevectors along $k_x/k_0 = 0$ and $k_y/k_0 = 0$ (shown as rectangles on the FP image). The resulting light was analyzed for its output polarization. (a) and (b) are output polarization resolved spectra along $k_x/k_0$ and $k_y/k_0$ axes respectively after filtering narrow range of wavevectors near $k_x/k_0 = 0$ using a narrow slit of width 200 µm. (c) and (d) are output polarization resolved spectra along $k_x/k_0$ and $k_y/k_0$ axes respectively after filtering narrow range of wavevectors near $k_y/k_0 = 0$ using a narrow slit of width 200 µm.

Fig. S11 shows the importance of momentum resolved spectroscopy. When a slit is introduced in the conjugate FP and then the polarization of the light is analyzed, we start to see hidden features in the polarization structure. Fig. S11 (a) and (b) shows the output polarization resolved spectrum, for slit placed along $k_x/k_0 = 0$ line, along $k_x/k_0$ axis and $k_y/k_0$ axis (0° and 90° analyzer positions) respectively. In this case, one can clearly see the difference of intensities in the split modes for two orthogonal analyzer positions. When the similar experiment was conducted for slit placed along $k_y/k_0 = 0$ line (fig. S11 (c) and (d)), we get exactly opposite signatures. Looking carefully at fig. S11 (a) and (c), one can easily notice that the polarization of the mode 'b' has been rotated by 90° and so is the polarization of mode 'a'. But they are orthogonal in nature. Over the full k-plane the polarization of the modes, then, will rotate 360°, making the modes azimuthally and radially poalrized.

# S12. Momentum resolved spectroscopy of strongly coupled microresonators

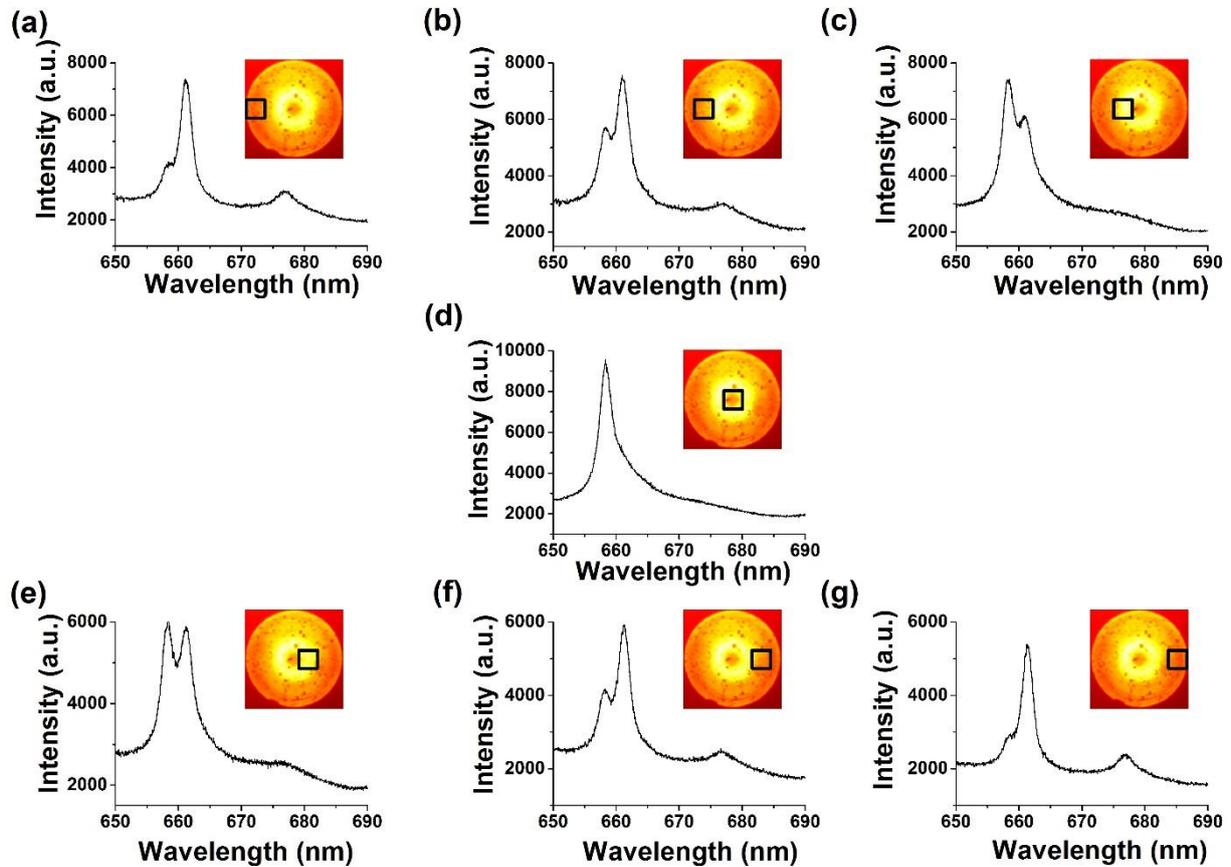

Figure S12: Momentum resolved spectra collected from the microsphere coupled to Au mirror system. The outcoupled light was filtered using a square pinhole of size 200 µm to understand spectral signatures at different k-points. (a) - (g) are k resolved spectra by placing pinhole at points along $k_y/k_0 = 0$ as shown in the insets.

Figures S12 (a)-(g) shows the k-resolved spectra of molecular fluorescence coupled to microsphere by placing a square aperture of size 200µm×200µm at the conjugate back focal plane. They clearly show the spectral distribution of split modes as a function of $k_x$, also outlines the fact that split modes occupy different wavevectors.

# S13. Calculated near field spectra from single microsphere

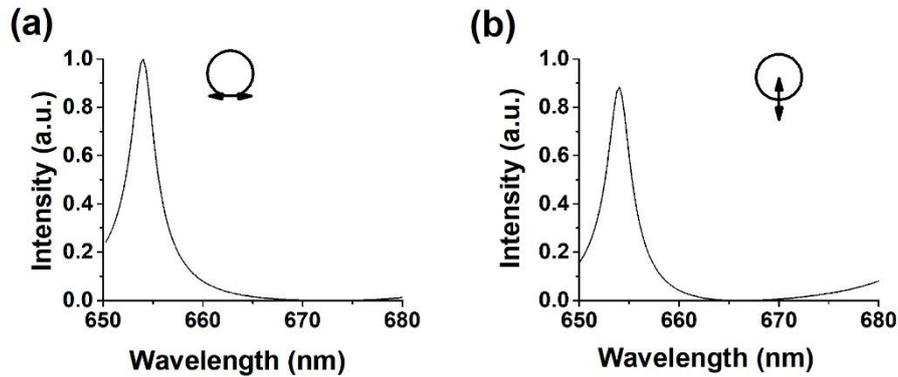

Figure S13: Calculated near field spectra by exciting an individual microsphere using dipoles oriented along (a) x and (y) axes respectively as shown in the insets.

# S14: Fourier analysis of microsphere coupled to Au film

For the microspheres with relatively lower $l$ values (large *distance from the substrate*), which is the present case, the electric field of WGMs will be concentrated near the equatorial plane. The local electric field of the mode is ring-shaped. Hence its projection onto the Fourier plane will also be a ring-like structure (see fig 2 (a) and (c)). The $k_x$ and $k_y$ values of any mode will then be intimately related to the distance from the substrate. Larger the *distance of substrate*, the ring in the FP will have large ($k_x,k_y$) values.

E-k spectrum gives information about the momentum distribution of the individual mie modes. Modes with degenerate |*m*| (-*l* to *l*) will span a range of wavevectors at a particular value of wavelength. This is seen in fig. 2 (d), where wavevector distribution of individual modes are shown. When the degeneracy is lifted by coupling it to its image sphere, we have ranges of '*m*' s associated with different wavelengths. Consider the case of peak 'a' and 'b' (fig. 1 (c)). Peak 'a' occurs at lower wavelength and hence associated with modes, *m*, that are nearer to the substrate. This is because of the fact that the modes with lower values of |*m*| have higher energy and hence lower wavelength. This makes mode 'a' emit at lower k values, as the electric field is located nearer to the substrate. The projection of such field onto the k-space will be near k=0. On the other hand peak 'b' contains modes away from the substrate. This makes split mode 'b' to emit at higher wavevectors when compared to its counterpart (see fig.2 (c)).

Also modes near to the substrate (modes with lower value of |*m*|) will be enhanced more than the modes near the equatorial plane [5]. And the modes near to the substrate project onto lower $k/k_0$ values. Hence for the microsphere on gold film, the majority of emission happens at lower k-values (see fig. 2(a)). This explains the intricate relationship between the mode numbers of the WGM and its wavevector signatures.